\documentclass[12pt, a4paper]{article}
\usepackage{amsmath}
\usepackage{amssymb}
\usepackage{amsfonts}

\usepackage{geometry}                
\usepackage{graphicx}
\usepackage{booktabs}
\usepackage{color}
\usepackage{natbib}
\usepackage{lscape}
\usepackage{rotating}
\usepackage{url}

\definecolor{purple}{RGB}{138,43,226}
\definecolor{bordeaux}{RGB}{139,0,0}
\definecolor{red}{RGB}{255,0,0}
\definecolor{grey}{RGB}{211,211,211}

\newcommand{\stanislav}{1}

\if0\stanislav
{
} \fi

\newcommand{\chapquote}[3]{\begin{quotation} \textit{#1} \end{quotation} \begin{flushright} - #2, \textit{#3}\end{flushright} }

\def\dsp{\def\baselinestretch{1.2}\large\normalsize}
\dsp
\title{Forecasting dynamic return distributions \\ based on ordered binary choice\thanks{We are grateful to the editor Dick van Dijk and two anonymous referees for useful comments and suggestions, which have greatly improved the article.
}}

\author{Stanislav {\sc Anatolyev}\thanks{CERGE-EI, Politick\'{y}ch v\v{e}z\v{n}\r{u} 7, 11121, Prague, Czech Republic; New Economic School, 45 Skolkovskoe Shosse, Moscow, 121353, Russia.}
\and
Jozef {\sc Barun\'{i}k}\thanks{Institute of Economic Studies, Charles University, Opletalova 26, 11000, Prague, Czech Republic; Institute of Information Theory and Automation, Academy of Sciences of the Czech Republic, Pod Vodarenskou Vezi 4, 18200, Prague, Czech Republic.}
}

\date{}

\begin{document}
\maketitle

\begin{abstract}

\noindent We present a simple approach to forecasting conditional probability distributions of asset returns. We work with a parsimonious specification of ordered binary choice regression that imposes a connection on sign predictability across different quantiles.
The model forecasts the future conditional probability distributions of returns quite precisely when using a past indicator and past volatility proxy as predictors. Direct benefits of the model are revealed in an empirical application to the 29 most liquid U.S. stocks. The forecast probability distribution is translated to significant economic gains in a simple trading strategy. Our approach can also be useful in many other applications where conditional distribution forecasts are desired.
\vspace{0.15cm}

\vspace{0.15cm}

\end{abstract}

\newpage
\section{Introduction}

\chapquote{``Those who have knowledge, don't predict. Those who predict, don't have knowledge."}{Lao Tzu}{(c. 604--531 B.C.)}

Several decades of research provides overwhelming evidence about the predictability of the first two moments of stock return distributions. Expected values of stock returns are predictable to some extent using economic variables \citep{keim1986predicting,fama1989business,ang2006stock,viceira2012bond}, while the conditional second moment can be well characterized by simple volatility models, or even measured from high frequency data \citep{bollerslev1986generalized,andersen2003modeling}. While volatility forecasting quickly became central to the financial econometrics literature due to its importance for risk measurement and management, research focusing on the entire return distribution still occupies a small fraction of the literature.\footnote{Few studies focused on directional forecasts, or threshold exceedances \citep{christoffersen2006financial,chung2007model,nyberg2011forecasting}.}

One of the main reasons why researchers may not be focused on characterizing the entire return distribution is a prevailing practice of convenient mean-variance analysis that is still central to modern asset pricing theories. Unfortunately, investor choices guided using the first two moments are restricted by binding assumptions, such as the multivariate normality of stock returns or a quadratic utility function. More importantly, an investor is restricted to have classical preferences based on von Neumann-Morgenstern expected utility. In contrast to this, \cite{rostek2010quantile} recently developed a notion of quantile maximization and quantile utility preferences. This important shift in decision-theoretic foundations provokes us to depart from the limited mean-variance thinking and work with entire distributions.

Specification and estimation of an entire conditional distribution of future price changes is useful for a number of important financial decisions. Prime examples include portfolio selection when returns are non-Gaussian, (tail) risk measurement and management, market timing strategies with precise entries and exits reflecting information in tails. Despite its importance, forecasting the conditional distribution of future returns has so far attracted little attention in contrast to point forecasts and its aforementioned uncertainty. In this article, we present a simple approach to forecasting a conditional distribution of stock returns using a parameterized ordered binary choice regression. Although focusing on financial returns, we note that our approach may be useful to many other applications where the conditional distribution forecasts are of interest.

The majority of studies focusing on prediction of conditional return distributions characterize the cumulative conditional distribution by a collection of conditional quantiles \citep{engle2004caviar,cenesizoglu2008distribution,vzikevs2014semi,pedersen2015predictable}. In contrast, in a notable contribution, \cite{foresi1995conditional} focus on a collection of conditional probabilities and  describe the cumulative distribution function of excess returns using a set of separate logistic regressions. To be able to approximate the distribution function, \cite{foresi1995conditional} estimate a sequence of conditional binary choice models over a grid of values corresponding to different points of the distribution. \cite{peracchi2002estimating} argues that the conditional distributions approach has numerous advantages over the conditional quantile approach, and \cite{leorato2015comparing} continue their comparison further. The approach has also been considered by \cite{fortin2011decomposition}, \cite{rothe2012partial}, \cite{chernozhukov2013inference}, \cite{hothorn2014conditional}, and \cite{taylor2016using}.

In this article, we further develop the ideas set forth by \cite{foresi1995conditional} and present a related simple model for forecasting conditional return distributions. The proposed model is based on an ordered binary choice regression, which is able to forecast the entire predictive distribution of stock returns using fewer parameters than the set of separate binary choice regressions. To achieve this substantial reduction in the degree of parameterization, we tie the coefficients of predictors via smooth dependence on corresponding probability levels. Our specification can be motivated in a semiparametric way, as we approximate smooth probability functions by low-order polynomials.

The probability forecasts are conditional on past information contained in returns, as well as their volatility proxy. The main reason for choosing volatility as one of the explanatory variables is that the cross-sectional relation between risk and expected returns, generally measuring a stock's risk as the covariance between its return and some factor, is well documented in the literature. In the laborious search for proper risk factors, volatility plays a central role in explaining expected stock returns for decades. Although predictions about expected returns are essential for understanding classical asset pricing, little is known about the potential of these factors to precisely identify extreme tail events of the return distribution.

In our illustrative empirical analysis, we estimate conditional distributions of the 29 most liquid U.S. stocks and compare their generated forecasts with those from the buy-and-hold strategy as well as several benchmarks -- a collection of separate binary choice models, a fully specified conditional density, and historical simulation. The benefits of our approach translate into significant economic gains in a simple trading strategy that uses conditional probability forecasts.

We provide the package \texttt{DistributionalForecasts.jl}
in the \textsf{Julia} software for estimating the model introduced in this article. The package is available at \url{https://github.com/barunik/DistributionalForecasts.jl}.

The article is organized as follows. Section 2 describes the model and emphasizes its differences with the collection of separate binary choice models. Section 3 contains information about the data we use and lays out details of particular specifications. In Section 4, empirical results are presented. Section 5 concludes. The Appendix contains more technical material and details of some procedures used in the empirical application.

\section{Model}

We consider a strictly stationary series of financial returns $r_{t}$, $t=1,\dots ,T$.
Our objective is to describe, as precisely as possible, the conditional cumulative return distribution $F\left( r_{t}\,|\,\mathcal{I}_{t-1}\right) ,$ where $\mathcal{I}_{t-1}$ includes the
history of $r_{t}$ as well as, possibly, past values of other observable variables.

Consider a partition of the support of returns by $p>1$ fixed cutoffs, or thresholds%
\begin{equation*}
c_{1}<c_{2}<\dots <c_{p},
\end{equation*}%
and define $c_{0}=-\infty $ and $c_{p+1}=+\infty $ for convenience.
The higher $p$, the more precise the description of the conditional distribution will be (see a full discussion later in the section).
The partition $\{c_{j}\}_{j=0}^{p+1}$ is arbitrary subject to the ordering restrictions. One intuitive partition corresponds to empirical quantiles of returns: each $c_{j}$ is an empirical $\alpha _{j}$-quantile of returns, $j=1,\dots ,p$, where $0<\alpha _{1}<\alpha _{2}<\dots <\alpha _{p}<1$ are $p$ probability levels; a reasonable grid for the probability levels is a regularly spaced unit interval $[0,1]$.
Alternatively, and perhaps more judiciously, the partition $\{c_{j}\}_{j=1}^{p}$ and thresholds $\{\alpha_{j}\}_{j=1}^{p}$ can be tied to some volatility measure to reflect the time-varying spread of returns due to the changing shape of the conditional distribution. Thus, in general the elements of the partition are time-varying and implicitly indexed by $t$.

Let $\Lambda :u\mapsto \left[ 0,1\right] $ be a (monotonically increasing) link
function. Both unordered and ordered binary choice models are represented by a collection of conditional probabilities
\begin{equation}
\Pr \{r_{t}\leq c_{j}|\mathcal{I}_{t-1}\}=\Lambda \left( \theta _{t,j}\right) ,\quad
j=1,\dots ,p,\label{Link}
\end{equation}%
for some specification of the driving processes $\theta _{t,j},$ $j=1,\dots ,p.$
For convenience, define $\Lambda \left( \theta _{t,0}\right) =0$ and $%
\Lambda \left( \theta _{t,p+1}\right) =1.$

Let $x_{t-1,j},$ $j=1,\dots ,p$ be a vector of predictors for $\mathbb{I}%
_{\{r_{t}\le c_{j}\}}$ that may depend on $j$ via the dependence of some of them on
$c_{j}$. For instance, one of predictors may be $\mathbb{I}%
_{\{r_{t-1}\le c_{j}\}},$ the past indicator (dependent on $j$), while another
may be $r_{t-1},$ the past return (independent of $j$), yet another may represent
some volatility measure (also non-specific to $j$). Suppose for
simplicity that the number of predictors in $x_{t-1,j}$ is the same for all $%
j$ and equals $k.$

In the unordered model, the specification for the underlying process $\theta _{t,j}$ is%
\begin{equation}
\theta _{t,j}=\delta _{0,j}+x_{t-1,j}^{\prime }\delta _{j}.\label{SLtheta}
\end{equation}%
There are no cross-quantile restrictions, and each binary choice problem is
parameterized separately. This results in a flexible but highly parameterized
specification. In the proposed ordered model, we place cross-quantile restrictions on the
parameters. In particular, the coefficients of predictors are tied via
smooth dependence on the probability levels; this leads to a substantial decrease
in the degree of parameterization.

In the unordered model, no monotonicity, in the language of \cite{foresi1995conditional}, holds in general. That is, $\Pr \{r_{t}\leq
c_{j-1}|\mathcal{I}_{t-1}\}$ may exceed $\Pr \{r_{t}\leq c_{j}|\mathcal{I}_{t-1}\}$ with
positive probability even though $c_{j-1}<c_{j}.$
In the proposed ordered model, the monotonicity property in-sample is imposed automatically by the specification of the ordered binary choice likelihood function. This may require artificial adjustments of the conditional distribution values at some thresholds. Out-of-sample, the monotonicity is also not guaranteed to hold, but similar artificial measures can be applied. One simple way is to shift that value of the conditional distribution that violates monotonicity at a particular threshold to its value at the previous threshold plus an additional small figure.
An alternative way to ensure both in-sample and out-of-sample predictability is via rearrangement \citep{chernozhukov2009rearrangement}.
Given the generally low predictability of conditional probabilities for returns (and hence their low variability compared to their mean), a share of observations that need such adjustments is expected to be low (see below for empirical evidence); therefore we give a preference to the former, simpler method.

The specification for the underlying process $\theta_{t,j}$ is%
\begin{equation}
\theta _{t,j}=\delta _{0,j}+x_{t-1,j}^{\prime }\delta \left( \alpha
_{j}\right) ,\label{OLtheta}
\end{equation}%
where $\delta \left( \alpha _{j}\right) $ are coefficients that are
functions of the probability level $\alpha _{j}$. Each probability-dependent slope coefficient vector is
specified as follows: $\delta \left( \alpha _{j}\right) =\left(
\delta _{1}\left( \alpha _{j}\right) ,\dots ,\delta _{k}\left( \alpha
_{j}\right) \right) ^{\prime },$ where for each $\ell =1,\dots ,k,$
\begin{equation}
\delta _{\ell }\left( \alpha _{j}\right) =\kappa _{0,\ell
}+\sum_{i=1}^{q_{\ell }}2^{i}(\alpha _{j}-0.5)^{i}\cdot \kappa _{i,\ell },\label{OLdelta}
\end{equation}%
and $q_{\ell }\leq p-1.$ Note that each intercept $\delta _{0,j}$ is $j$%
-specific and represents an `individual effect' for a particular
probability level, while the slopes $\delta $'s do not have index $j,$ i.e. they
depend on $j$ only via dependence on $\alpha _{j}$'s. The motivation behind
such specification is semiparametric: any smooth function on $\left[ 0,1%
\right] $ can be approximated to a desired degree of precision by the system
of basis polynomials $\{\alpha _{j}-0.5,(\alpha _{j}-0.5)^{2},\dots ,(\alpha
_{j}-0.5)^{q}\}$ by making $q$ big enough. Because all $\alpha _{j}\in (0,1),
$ the polynomial form behaves nicely even for a large $q$; the basis
polynomials are uniformly bounded on $\left[ 0,1\right] .$ The additional
weights $2^{i}$ are introduced to line up the coefficients $\kappa _{i}$ on
a more comparable level.

Let us compare the degrees of parameterization of the unordered and ordered binary choice models.
Denote
\begin{equation*}
q=\sum_{\ell =1}^{k}q_{\ell }.
\end{equation*}%
In the unordered model, the total number of parameters is%
\begin{equation*}
K_{UO}=\left( 1+k\right) p
\end{equation*}%
(namely, one intercept $\delta _{0,j}$ and $k$ slopes $\delta _{j}$ in each
of $p$ equations for $\theta _{j}$), while in the ordered model, the total
number of parameters is%
\begin{equation*}
K_{O}=p+k+q
\end{equation*}%
(namely, $p$ intercepts $\delta _{0,j}$ and $k$ slopes $\delta \left( \alpha
_{j}\right) $, each parameterized via $1+q_{\ell }$ parameters). The
difference%
\begin{equation*}
K_{UO}-K_{O}=k\left( p-1\right) -q
\end{equation*}%
is larger the larger $p$ is, the fineness of the partition by thresholds. The
resulting difference is also positively related to the number of predictors
used.\footnote{%
In our empirical illustration, $p=37,$ $k=2$ and $q_{1}=2,$ $q_{2}=3.$
Hence, $K_{UO}=111$ while $K_{O}=44,$ and the difference is $%
K_{UO}-K_{O}=67. $}

In the unordered model, the composite loglikelihood corresponding to observation $t$ is%
\begin{equation}
\ell _{t}^{UO}=\sum_{j=1}^{p+1}\mathbb{I}_{\{r_{t}\leq c_{j}\}}\ln \left(
\Lambda \left( \theta _{t,j}\right) \right) ,\label{SLtotalL}
\end{equation}%
and the total composite likelihood $\sum_{t=1}^{T}\ell _{t}^{UO}$ can be split into $p$
independent likelihoods $\sum_{t=1}^{T}\ell _{t}^{(j)},$ where
\begin{equation}
\ell _{t}^{(j)}=\mathbb{I}_{\{r_{t}\leq c_{j}\}}\ln \left( \Lambda \left(
\theta _{t,j}\right) \right) ,\label{SLindL}
\end{equation}%
to be maximized over the parameter vector $\left( \delta _{0,j},\delta
_{j}\right) ^{\prime }.$ In the ordered model, the loglikelihood
corresponding to observation $t$ is%
\begin{equation}
\ell _{t}^{O}=\sum_{j=1}^{p+1}\mathbb{I}_{\{c_{j-1}<r_{t}\leq c_{j}\}}\ln
\left(\Delta_j\Lambda_t\right) ,\label{OLtotalL}
\end{equation}%
where $\Delta_1\Lambda_t=\Lambda \left( \theta _{t,1}\right),$
$\Delta_j\Lambda_t=
\Lambda \left( \theta _{t,j}\right) -\Lambda \left( \theta
_{t,j-1}\right) $
for $j=2,\dots ,p$, and $\Delta_{p+1}\Lambda_t=1 -\Lambda \left( \theta
_{t,p}\right).$
The total likelihood $\sum_{t=1}^{T}\ell _{t}^{O}$ is to be maximized
over the parameter vectors $\left( \delta _{0,1},\dots ,\delta _{0,p}\right)
^{\prime }$ and $\left( \kappa _{0,1},\dots ,\kappa _{0,k},\kappa
_{1,1},\dots ,\kappa _{q_{k},k}\right) ^{\prime }$. Under mild suitable conditions, the estimates of the parameter vector are expected to be consistent for their pseudotrue values and asymptotically normal around them, with a familiar sandwich form of the asymptotic variance.

Of course, each difference $\Delta_j\Lambda_t $ needs to be positive. This monotonicity
property, though not guaranteed, is easier to enforce in maximization of the joint ordered likelihood $\sum_{t=1}^{T}\ell _{t}^{O}$ than in separate maximizations of independent unordered likelihoods, as the common parameters will automatically adjust to these monotonicity restrictions.
However, if the degrees of freedom are insufficient to ensure this for all predictor values in the sample, to prevent (rare) realizations of negative differences for particular $t,$
monotonicity can be enforced by imposing constraints $\Delta_j\Lambda_t \geq \varepsilon
$ for all \thinspace $j$ and $t,$ where $\varepsilon $ is some small number.%
\footnote{%
In our empirical illustration, we made each difference to be bounded below
by $\varepsilon =10^{-6}$ by force. Such strategy has resulted in only less
than 1\% of such interferences among all differences during in-sample estimation and less than 2\% during out-of-sample forecasting.}

Computationally, it is convenient to maximize the total likelihood in a number of steps. The reason is that an arbitrary initial parameter vector is likely to result in incomputable likelihood because of numerous violations of the monotonicity property. The idea is to first determine approximate values of individual intercepts and slopes, subject to their monotonicity, then relaxing the restrictions on the slopes using the evaluated values as starting points for corresponding parts of the parameter vector. Towards this end, we propose and further use the following algorithm\footnote{The estimation can be done using the package \texttt{DistributionalForecasts.jl}
developed by the authors in the \textsf{Julia} software. The package is available at
\url{https://github.com/barunik/DistributionalForecasts.jl}
}:
\begin{enumerate}
\item[Step 1.]Run a series of separate binary choice models (\ref{Link}) and (\ref{SLtheta}) with the specification $\theta _{t,j}=\delta _{0,j}+x_{t-1,j}^{\prime }\delta_j$ with $\delta_j =\left(\delta_{j,1},\dots ,\delta_{j,k} \right) ^{\prime },$ $j=1,\dots ,p$, by maximizing the individual likelihoods (\ref{SLindL}), and call the obtained estimates~$\bar\delta _{0,j}$ and $\bar \delta_j$.
\item[Step 2.]For each $\ell=1,\dots ,k$, run a linear regression $\bar \delta _{\ell } =\kappa _{0,\ell}+\sum_{i=1}^{q_{\ell }}2^{i}(\alpha-0.5)^{i}\cdot \kappa _{i,\ell }$, where $\delta _{\ell }=\left(\delta_{1,\ell},\dots ,\delta_{p,\ell} \right) ^{\prime }$ and $\alpha = \left( \alpha_1,\dots \alpha_p\right)^{\prime }$, and call the obtained estimates $\bar \kappa_{i,\ell}$, $i=0,1,\dots ,q_{\ell}$.
\item[Step 3.]Run the ordered binary choice model (\ref{Link}) and (\ref{OLtheta}) with specification (\ref{OLdelta}) by maximizing the total likelihood (\ref{OLtotalL}) using $\bar\delta _{0,j}$ as starting points for $\delta _{0,j}$, $j=1,\dots ,p$, and $\bar \kappa_{i,\ell}$  as starting points for $\kappa _{i,\ell},$ $i=0,1,\dots ,q_{\ell}$, $\ell=1,\dots ,k.$
\end{enumerate}

Having estimated the conditional return distribution evaluated at the threshold values, one can obtain the entire (continuous) conditional distribution by using interpolation schemes that preserve monotonicity of the outcome. Towards this end, we apply the Fritsch--Carlson monotonic cubic interpolation \citep{fritsch1980interpolation} (see Appendices A.1 and A.2) and use the result for testing the quality of the estimated distribution (see Appendices A.3 and A.4).

The quality of approximation of the constructed conditional distribution in the ordered model is determined by a number of factors, the precision of interpolation being the least important. There is an important tradeoff between the number of thresholds (and hence the precision of the interpolation input), and the degree of parameterization and hence the amount of estimation noise. Yet another factor is the flexibility of specification of the slopes $\delta $'s on $\alpha _{j}$'s. It seems reasonable to set the system of thresholds fine enough (as long as one does not come close to the computability  limits) to describe the distribution precisely enough but not too fine so that there are a sufficient number of observations that fall between each pair of adjacent thresholds. One may also afford higher flexibility to the slope specification for larger sample sizes; however, low numbers are usually pretty adequate in semiparametric setups in practice. One may also employ formal model selection criteria such as the Bayesian information criterion to choose the optimal orders of polynomials in slope specifications.

\section{Data and Empirical Specification}

We study the conditional distribution forecasts of 29 U.S. stocks\footnote{Apple Inc. (AAPL), Amazon.com, Inc. (AMZN), Bank of America Corp (BAC), Comcast Corporation (CMCSA), Cisco Systems, Inc. (CSCO), Chevron Corporation (CVX), Citigroup Inc. (C), Walt Disney Co (DIS), General Electric Company (GE), Home Depot Inc. (HD), International Business Machines Corp. (IBM), Intel Corporation (INTC), Johnson \& Johnson (JNJ), JPMorgan Chase \& Co. (JPM), The Coca-Cola Co (KO), McDonald's Corporation (MCD), Merck \& Co., Inc. (MRK),Microsoft Corporation (MSFT), Oracle Corporation (ORCL), PepsiCo, Inc. (PEP), Pfizer Inc. (PFE), Procter \& Gamble Co (PG), QUALCOMM, Inc. (QCOM), Schlumberger Limited. (SLB), AT\&T Inc. (T), Verizon Communications Inc. (VZ), Wells Fargo \& Co (WFC), Wal-Mart Stores, Inc. (WMT), Exxon Mobil Corporation (XOM).} that are traded on the New York Stock Exchange. These stocks have been chosen according to market capitalization and their liquidity. The sample under investigation spans from August 19, 2004 to December 31, 2015. We consider trades executed during U.S. business hours (9:30--16:00 EST). In order to ensure sufficient liquidity and eliminate possible bias we explicitly exclude weekends and bank holidays (Christmas, New Year's Day, Thanksgiving Day, Independence Day). In total, our final dataset consists of 2826 trading days, 500 of which are used for in-sample estimation, and the rest 2326 for out-of-sample forecasting using a rolling window scheme with the window size of 500 days. We split the sample to have a much larger out-of-sample portion as we perform an extensive set of tests, robustness checks and inter-model comparisons on it.

Next we give the details of our empirical specification. For both ordered and unordered models,
we use the logit link function%
\begin{equation*}
\Lambda \left( u\right) =\frac{\exp (u)}{1+\exp (u)}
\end{equation*}%
resulting in logit specifications. We consider a partition of the return space into 37 equally spaced probability levels\footnote{The use of more extreme values of probability levels such as $1\%$ and $99\%$ leads, with our in-sample window of 500 observations, to large estimation uncertainty in the tails, but ends up successful when the model is estimated on the whole sample.} from $\alpha \in (5\%,95\%)$, i.e., a grid with step $2.5\%$ resulting in $p=37$ quantiles in total.\footnote{We have also used partitions with $p=19$ and $p=73$ equally spaced probability levels; for details, see subsection \ref{sensitivity}.} We use a time-varying partition that changes with the rolling window. In every window, $c_j$ is computed as $\gamma(\alpha_j)\sqrt{\sigma_t^2}$, where $\gamma(\alpha_j)$ is a quantile of the standard normal distribution, and $\sigma_t^2$ is a conditional variance of returns in the corresponding window computed from the RiskMetrics of JPMorgan Chase standards as an exponentially weighted moving average with a decay factor of 0.94.


We choose $k=2$ and the predictors to be
\begin{equation*}
x_{t-1,j}=\binom{\mathbb{I}_{\{r_{t-1}\leq c_{j}\}}}{\ln \left( 1+\left\vert r_{t-1}\right\vert \right) }
\end{equation*}%
for all $j=1,\dots ,p$. The first predictor is a lagged indicator corresponding to the probability level $%
\alpha _{j},$ which is plausibly supposed to have the highest predictability power among all such indicators. The second predictor is a proxy for a volatility measure,
with the absolute return dampened by the logarithmic
transformation. Note that the first predictor is specific to a specific quantile,
while the second predictor is common for all quantiles. In principle, one could specify
all predictors to be the same across the quantiles, or, on the opposite, all predictors may vary with the quantile.
In the ordered model, we set, after some experimentation with statistical significance of higher-order polynomials,
$q_{1}=2$ and $q_{2}=3.$\footnote{We also perform model selection analysis with the Bayesian information criterion in the ordered model; for details, see subsection \ref{sensitivity}.} That is, the polynomial is quadratic in the probability level
$\alpha $ for the past indicator and cubic for the past volatility proxy.

The full specification of the model for $j=1,\dots ,p$ empirical quantiles is
\begin{equation*}
\Pr \{r_{t}\leq c_{j}|\mathcal{I}_{t-1}\}=\frac{\exp (\theta _{t,j})}{1+\exp (\theta _{t,j})},
\end{equation*}%
\begin{equation*}
\theta _{t,j}=\delta _{0,j}+\delta_1 \left( \alpha
_{j}\right) \mathbb{I}_{\{r_{t-1}\leq c_{j}\}} +\delta_2 \left( \alpha
_{j}\right) \ln \left( 1+\left\vert r_{t-1}\right\vert \right),
\end{equation*}%
with the coefficient functions
\begin{equation*}
\delta_1 \left( \alpha _{j}\right) =\kappa _{0,1}+
2 (\alpha _{j}-0.5) \cdot \kappa _{1,1 }
+2^2 (\alpha _{j}-0.5)^2 \cdot \kappa _{2,1 },
\end{equation*}%
\begin{equation*}
\delta _2\left( \alpha _{j}\right) =\kappa _{0,2}
+2(\alpha _{j}-0.5)\cdot \kappa _{1,2}
+2^{2}(\alpha _{j}-0.5)^{2}\cdot \kappa _{2,2}
+2^{3}(\alpha _{j}-0.5)^{3}\cdot \kappa _{3,2}.
\end{equation*}%

In total, there are $K_{O}=44$ parameters: $p=37$ individual intercepts $\delta _{0,j}$ and $k=2$ slopes $\delta \left( \alpha_{j}\right) $, one parameterized via $1+q_{1}=3$ parameters, the other via $1+q_{2}=4$ parameters.
This parametrization is parsimonious enough and approximates the distribution quite well, and additional terms do not bring significant improvements. Hence, estimating seven parameters $\kappa_{i,\ell}$ in addition to individual intercepts is enough to approximate the conditional return distribution.

\section{Empirical Findings}

We now present the results of estimating the conditional distribution function of returns using the ordered binary choice model. We consider forecasts for 29 stocks, so first we present individual estimates of three illustrative stocks, namely Intel Corporation (INTC), QUALCOMM, Inc. (QCOM), and Exxon Mobil Corporation (XOM), and then we present the results for all 29 stocks in an aggregate form.

After presenting the parameter estimates, we evaluate the statistical as well as economic significance of the predicted distributions. Further, we compare the performance of the ordered model with popular and challenging benchmarks used in the literature. Namely, we include two candidate models that perform best \citep{kuester2006var} -- an asymmetric generalized autoregressive conditional heteroscedastic model with skewed-t distribution (GARCH henceforth) and GARCH-filtered historical simulation (FHS henceforth). In implementing both alternative methods, we follow \cite{kuester2006var}.

\subsection{Parameter estimates}
Table 1 
shows estimates of cutoff-specific intercepts $\delta _{0,j}$ in the ordered logit model for the three illustrative stocks. The intercepts have expected signs -- negative for probability levels to the left of 50\% and positive to the right of 50\% -- due to monotonically increasing $\Lambda(\cdot)$ and low predictability of predictors and exhibit expected monotonic behavior increasing from the left tail to the right tail, thus generating an increasing cumulative distribution function (assuming zero predictors). The intercepts are statistically significant in most cases except for few a cutoff points near the median. The intercept values are quite similar across three stocks, though there is some variability.

The left plot shown in Figure \ref{fig:pars2} collects the intercept estimates for all 29 stocks and reveals that indeed the intercepts are similar across the stocks. The plot has a shape similar to an inverse logit link function $\Lambda \left( u\right)$ defined earlier, with a stronger effect in the tails. A flexible $j$-specific intercept allows for controlling individual quantile effects, and shows that unconditional expectations are an important part of the predicted distribution.

The estimates of the seven coefficients $\kappa _{i,\ell }$ driving the slopes on the predictors
for the three illustrative stocks, are shown in Table 2. The coefficients $\kappa _{i,1}$ correspond to the lagged indicator predictor $\mathbb{I}_{\{r_{t-1}\leq c_{j}\}}$. All parameter estimates for the lagged indicator are of a small magnitude, and some are highly statistically significant. The left plot in Figure \ref{fig:pars1} complements these findings with estimates of the lagged indicator coefficients for all 29 stocks shown by the box-and-whisker plots. The estimates document a heterogeneous effect of the lagged indicator on the future probabilities for different stocks. While we document zero coefficients for some stocks, the quadratic term seems to play a big role in others.

The second predictor, the past volatility proxy $\ln \left( 1+\left\vert r_{t-1}\right\vert \right)$, carries even more important information about future probabilities.
The estimated coefficients $\kappa _{i,2}$ corresponding to the volatility predictor  reveal that
the cubic polynomial has many statistically significant coefficients (see Table 2). The right plot of Figure \ref{fig:pars1} shows estimated coefficients for all 29 stocks as box-whisker plots. We can see that $\kappa _{1,2}$ and $\kappa _{3,2}$ are significantly different from zero for most of the stocks, and so the past volatility proxy contributes strongly to the prediction of return distributions.

Figure \ref{fig:pars2} plots the functions $\delta_1 \left(\alpha _{j}\right)$, and $\delta_2 \left(\alpha _{j}\right)$, in addition to the intercepts. Corresponding to heterogenous parameters $\kappa _{i,1}$, the function $\delta_1 \left(\alpha _{j}\right)$ is also heterogeneous for 29 stocks, exhibiting a mixed effect of the lagged indicator predictor (shown in the middle plot of Figure \ref{fig:pars2}). Overall, we can see that the effect is small. Coefficient function $\delta_2 \left(\alpha _{j}\right)$ implied by the past volatility proxy shows a similar impact.

Figure \ref{fig:CDFplot} depicts an interpolated predictive conditional CDF of returns on the interval $[-1.5,1.5]$ for an arbitrary stock\footnote{The minimal and maximal values of returns in the whole sample for this stock are approximately $-0.20$ and $0.20$.} evaluated at arbitrary 100 out-of-sample periods. One can observe how the cumulative distribution varies over time. There is a certain distribution clustering, i.e. the CDF possesses some persistence, while at some periods the CDF shape stands out from the cluster.

\subsection{Statistical Fit}

To assess the adequacy of the predicted distribution, we run the generalized autocontours specification tests of \cite{gonzalez2015contours}. This test verifies whether the collection of out-of-sample generalized residuals (also known as probability integral transform) together with their lags are scattered uniformly inside a hypercube of corresponding dimension; see Appendix A.3 for a detailed description. We use lag $L=1$ in the contour-aggregated test with a simple collection of sides $\alpha=(0.25,0.50,0.75)'$ and a larger collection coinciding with our full grid
$\alpha=(\alpha _{1},\alpha _{2},\dots ,\alpha _{p})'.$ We use side $\bar\alpha=0.5$ in the lag-aggregated test with $L=3$ and $L=10$.
Figure \ref{fig:testplot} shows the distributions of p-values from the \cite{gonzalez2015contours} tests for the ordered logit model for all 29 stocks. All four variations of the test -- two contour-aggregated and two lag-aggregated -- show the adequacy of the estimated conditional distribution.\footnote{We did not account for the estimation noise in constructing the test. Because it generally tends to increase asymptotic variances, the p-values would be even higher if the estimation noise was accounted for.}

We also compare probabilistic forecasts of different models in terms of proper scoring rules \citep{gneiting2007scoring}, namely the Brier score and the continuous ranked probability score  (CRPS); see Appendix A.4 for a detailed description.\footnote{We follow \cite{gneiting2007scoring} and define the CRPS with a minus sign so that its larger values are preferred to smaller values.}
Figure \ref{fig:scoresplot} shows the average score values for the four models (ordered logit, collection of separate logits, GARCH and FHS). All four approaches deliver Brier score that have similar median values though differing a bit in dispersion. The CRPS scores are very similar across all four approaches, though the logit models marginally dominate.

\subsection{Economic Significance}

To understand the economic usefulness of the proposed model, we study a simple profit rule for timing the market based on the model forecasts. The idea is to explore information from the entire distribution. To build a trading strategy, we explore the difference between the predicted conditional probability and unconditional probability, which indicates if positive or negative returns are predicted with higher probability. In the spirit of the previous literature \citep{breen1989economic,pesaran1995predictability,anatolyev2010modeling}, we evaluate the model forecasts in terms of profits from a trading strategy for active portfolio allocation between holding a stock and investing in a risk-free asset. The detailed construction of the trading strategy is described in Appendix A.5.

Table 3 
summarizes the results of running the trading strategies in terms of a mean and median return, volatility, as well as Sharpe ratio. While the benchmark market strategy earns a 0.806 return with  a 0.244 volatility, GARCH and FHS generate similar returns with a lower volatility. When an investor uses predictions about the entire distribution from separate logit models, an improved mean return of 1.088 is obtained with a much lower volatility. Finally, our proposed ordered logit model generates a 1.296 mean return with a volatility similar to that of separate logits. The Sharpe ratio showing a risk-adjusted return, or a average return per unit of volatility, reveals that the ordered logit model generates the highest returns when one takes risk into account. The figures for median values confirm this result.

Figure \ref{fig:tradingplot} shows returns, volatilities, and Sharpe ratios for all 29 stocks using box-and-whisker plots. The figure reveals that the naive buy and hold strategy yields low returns with the highest volatility, which are moreover quite heterogeneous for all the considered stocks. GARCH and FHS improve the volatility estimates, hence yielding similar returns with a lower risk. In terms of risk-adjusted Sharpe ratios, the separate logistic regressions yield a similar result, while the ordered logit model makes a marked improvement. One can see positive Sharpe ratios for almost all considered stocks pointing to the highest risk-adjusted returns offered by the ordered model.

Figure \ref{fig:exthree} looks closely at the cumulative returns and drawdowns of the three illustrative stocks we used earlier. It can be seen that returns from the ordered logit strategy are consistent over time, with the lowest drawdown. This holds even in the case of a XOM that has been growing for the whole period, hence making it difficult to beat the buy and hold strategy.

Finally, we compare the relative performance of all five strategies in Figure \ref{fig:relative}. The top left plot in the Figure compares the ordered logit with the `market,' or naive buy and hold strategy, while the top right plot compares the ordered logit with separate logits. The plots at the bottom of Figure \ref{fig:relative} compare the ordered logit versus GARCH and FHS. We document a consistently better performance for the proposed ordered logit model in comparison with both the unordered logit as well as benchmark strategies for all 29 stocks.

\subsection{Sensitivity Analysis}
\label{sensitivity}

As noted before, we have chosen, after some experimentation, specification with $p=37$ thresholds and polynomial orders $q_{1}=2$ and $q_{2}=3.$ Here we report on the results with different specifications. As a robustness check, we do an investigation of the impact of partition fineness and polynomial complexity on the performance of the proposed method.

We have estimated and evaluated the model with different choices of partitions, namely $p=19$ (corresponding to twice as coarse a partition) and $p=73$ (corresponding to twice as fine a partition). On the one hand, $p=19$ may seem too coarse for the CDF approximation to be considered good. On the other hand, too high $p$ may seriously impact the model complexity: while the number of parameters for the basic partition $p=37$ equals $K_{O}=44,$ for the finer partition it equals 80, which is obviously too many for our out-of-sample exercise with 500 rolling in-sample observations.
In our experiments with $p=19$ and $p=73$,\footnote{The detailed results are available upon request from the authors.} some of the generalized autocontours tests (in particular, the three-side contour-aggregated test) exhibits the tendency to reject the constructed conditional distribution.
The results of economic evaluation though are not sensitive to the choice of partition, and the dominance of the proposed method based on the ordered logit over all the other methods still prevails.

The polynomial orders $(q_1,q_2)$ do not affect the degree of parsimony drastically, on the one hand, and on the other, the dependence of predictability on the probability level is presumably not very sophisticated to require higher-order powers. Hence, one would not expect high sensitivity to the choice of orders. We have run the proposed model with various combinations of polynomial orders $(q_1,q_2)$ around the running $(q_1,q_2)=(2,3)$ combination, and computed the value of the Bayesian information criterion (BIC) for each. The pattern of BIC is presented in Figure \ref{fig:sensitivityplot}. While the values of $q_1$ and $q_2$ such as 0 and 1 are obviously too small to capture the differences in predictability across the probability levels, there is little sensitivity once the orders reach the selected combination, which is clearly preferred by BIC en masse across the 29 stocks.

\section{Conclusion}

This article investigates the predictability of stock market return distributions. We propose a relatively parsimonious parametrization of an ordered binary choice regression that forecasts well the conditional probability distribution of returns. We subject the proposed model to a number of statistical tests for adequacy and to comparison with alternative methods. In order to see how useful the model is economically, we use distributional predictions in a simple market timing strategy. Using 29 liquid U.S. stocks, we find significant economic gains when compared to the benchmarks.

Our findings are useful for risk management and measurement or building trading strategies using the entire conditional distribution of returns. The model nevertheless has a much wider potential use in any application that exploits distribution forecasts, including forecasting of interest rates, term structures, and macroeconomic variables.
\newline
\newpage

\bibliographystyle{chicago}
\bibliography{references}

\newpage
\appendix
\section{Appendix}

\subsection{CDF interpolation}

The Fritsch--Carlson monotonic cubic interpolation \citep{fritsch1980interpolation}
provides a monotonically increasing CDF with range $[0,1]$ when applied to CDF estimates on a finite grid.

Suppose we have CDF $F(r)$ defined at points $(r_k,F(r_k))$ for $k=1,\dots ,K,$ where $F(r_0)=0$ and $F(r_K)=1$.
We presume that $r_k<r_{k+1}$ and $F(r_k)<F(r_{k+1})$ for all $k=0,\dots ,K-1,$ which is warranted by the continuity of returns and construction of the estimated distribution. First, we compute the slopes of the secant lines as $\Delta_k =(F(r_{k+1})-F(r_k)))/(r_{k+1}-r_k)$ for $k=1,\dots,K-1,$ and then the tangents at every data point as $m_1 = \Delta_1$, $m_k = \frac{1}{2}(\Delta_{k-1}+\Delta_k)$ for $k=2,\dots,K-1$, and $m_K = \Delta_{K-1}.$
Let $\alpha_k = m_k/\Delta_k$ and $\beta_k = m_{k+1}/\Delta_k$ for $k=1,\dots,K-1$.
If $\alpha_k^2 + \beta_k^2 > 9$ for some $k=1,\dots,K-1,$ then we set $m_k = \tau_k \alpha_k \Delta_k$ and $m_{k+1} = \tau_k \beta_k \Delta_k,$ with $\tau_k = 3(\alpha_k^2 + \beta_k^2)^{-1/2}$.
Finally, the cubic Hermite spline is applied: for any $r\in [r_k,r_{k+1}]$ for some $k=0,\dots,K-1,$
we evaluate $F(r)$ as
$$F(r) = (2t^3-3t^2+1)F(r_k)  + (t^3-2t^2+t)h r_{k}  + (-2t^3+3t^2) F(r_{k+1}) + (t^3-t^2)h m_{k+1},$$ where $h=r_{k+1}-r_{k}$ and $t=(r-r_{k})/h.$

\subsection{Generalized residuals}

The CDF specification testing is based on the properties of the generalized residuals (also known as probability integral transform).
First, for each out-of-sample period $t=R+1,\dots ,T,$ we apply the CDF interpolation algorithm with input data $(2r_{\min},0),(c_{j}, \widehat{\Pr} \{r_{t}\leq c_{j}|\mathcal{I}_{t-1}\}),(2r_{\max},1)$ for $j=1,\dots ,p,$ where $r_{\min}$ and $r_{\max}$ are minimal and maximal sample values of returns within the estimation portion of the sample. That is, we approximate the conditional CDF values outside the interval $[2r_{\min},2r_{\max}]$ by exact zero or exact one, which is reasonable, as the probability of such returns is negligible.
The generalized residual $\varepsilon_t ,$ $t=R+1,\dots ,T,$ is computed simply as an interpolated conditional CDF evaluated at $r_{t}.$

\subsection{CDF Testing}

The generalized residuals $\varepsilon_t $ have a familiar property that $\varepsilon_t \sim i.i.d.\,U[0,1]$. The univariate version of the generalized autocontours test of \cite{gonzalez2015contours} verifies whether the collection of $k$ out-of-sample generalized residuals together with their lags are scattered uniformly inside the $[0,1]^{2k}$ hypercube.

The testing procedure contains the following steps. Let the vector $\alpha$ contain $p_{\alpha}$ `sides' $\alpha_i\in (0,1]$, and consider pairs $(\varepsilon_t,\varepsilon_{t-\ell})$ of out-of-sample generalized residuals and their $\ell^{th}$ lags, $\ell=1,2,\dots ,L,$
$t=R+1,\dots ,T.$ Under the null hypothesis of correct specification when $\varepsilon_t \sim i.i.d.\,U[0,1]$, each side $\alpha_i$ is consistently estimated by the sample proportion of pairs $(\varepsilon_t,\varepsilon_{t-\ell}) $ falling into the corresponding generalized autocontour -- the subhypercube $\text{G-ACR}_{\alpha,\ell}=\times_{i=1}^{p_{\alpha}}[0,\sqrt{\alpha_i}]^2$:
$$\hat\alpha_{\alpha,\ell}=\frac{1}{T-R-\ell}\sum_{t=R+1+\ell}^{T}{\mathbb I}_{\{(\varepsilon_t,\varepsilon_{t-\ell})\in \text{G-ACR}_{\alpha,\ell}\}}.$$

The Gonz\'{a}lez-Rivera--Sun test exists in two chi-squared variations: contour-aggregated and lag-aggregated.
The contour-aggregated statistic gathers information from estimated generalized autocontours for a collection of different sides, keeping the lag, say $\bar\ell$, fixed. Let $\hat\alpha_{\bar\ell}=(\hat\alpha_{1,\bar\ell},\dots ,\hat\alpha_{p,\bar\ell})'.$
The lag-aggregated statistic gathers information from estimated generalized autocontours for a collection of different lags, keeping the side, say $\bar\alpha$, fixed. Let $\hat\alpha_{\bar\alpha}=(\hat\alpha_{\bar\alpha,1},\dots ,\hat\alpha_{\bar\alpha,L})'.$

Then, under the null of correct distributional specification,
$$
GRS_{\alpha,\bar\ell}=(\hat\alpha_{\bar\ell}-\alpha)'A_{\alpha,\bar\ell}^{-1}(\hat\alpha_{\bar\ell}-\alpha)\rightarrow^d \chi_{p_{\alpha}}^2
$$
and
$$
GRS_{\bar\alpha,L}=(\hat\alpha_{\bar\alpha}-\bar\alpha \iota_L)' A_{\bar\alpha,L}^{-1}(\hat\alpha_{\bar\alpha}-\bar\alpha \iota_L)\rightarrow^d \chi_L^2,
$$
where the matrices $A_{\alpha,\bar\ell}$ and $A_{\bar\alpha,L}$ contain asymptotic variances and covariances of elements of the estimated generalized autocontours, which are functions of elements of the vector $\alpha$ only and need not be estimated (see \cite{gonzalez2015contours} for more details), and $\iota_L$ is a column vector of ones of length $L$.

The rejection by the Gonz\'{a}lez-Rivera--Sun tests means that the generalized residuals are not likely to be uniform on [0,1] and/or fail to be serially independent.

\subsection{Scoring rules}

\cite{gneiting2007scoring} list several scoring rules that can be used to compare probabilistic forecasts of different models of (conditional) distributions.
The Brier score for the forecast at $t$ made at $t-1$ is
$$
B_t=-\sum_{j=1}^{p+1}\left(\mathbb{I}_{\{c_{j-1}<r_{t}\leq c_{j}\}}-\widehat{\Pr}{\{c_{j-1}<r_{t}\leq c_{j}\}} \right)^2,
$$
and is a quadratic criterion of deviations of binary realizations from probability forecasts.
The CRPS for the forecast at $t$ made at $t-1$ is
$$
CRPS_t=-\int_{-\infty}^{\infty}\left(
\widehat{\Pr} \{r_t\leq r|\mathcal{I}_{t-1}\}-\mathbb{I}_{\{r_t\leq r\}}\right)^2{\rm d}r,
$$
where the conditional CDF $\widehat{\Pr} \{r_t\leq r|\mathcal{I}_{t-1}\}$ is obtained by CDF interpolation (see Appendix A.1), while the integral is computed numerically using the Gauss-Chebyshev quadrature formulas (\cite{judd1998}, section 7.2) with 300 Chebychev quadrature nodes on
$\left[2r_{\min},2r_{\max}\right]$.

The average Brier score and average CRPS are computed by averaging $B_t$ and $CRPS_t$ over out-of-sample periods $t=R+1,\dots ,T.$

\subsection{Trading strategy}

To build a trading strategy, we use a simple rule exploring the difference between the predicted conditional probability and unconditional probability $\Pr\{r_t \le c_j\} = \alpha_j$. We sum the differences over the interval of empirical quantiles $[a,b]$ as
\begin{equation*}
S_t = \sum_{c_{j}=a}^b\left(\widehat{\Pr} \{r_{t}\leq c_{j}|\mathcal{I}_{t-1}\} - \alpha_j \right).
\end{equation*}%
If we sum all $p$ available quantiles, we are using the information from the entire distribution. If we want to use the information only about positive returns, we sum only half of the available empirical quantiles corresponding to the cutoffs at positive returns. For example, in case the positive returns are predicted with a higher probability than the negative returns for all corresponding empirical quantiles, the sum $S_t$ will be positive. Further, it may be useful to compare $S_t$ computed for the empirical quantiles corresponding to both negative and positive returns. After some experimentation, we obtain threshold values for each stock depending on the shape of the conditional distributions, generating consistent profits. Hence we build the trading strategy on $S_t$ exceeding these thresholds, but we note that this could further be optimized for maximum profits. In our setup, we use all quantiles, hence $a=c_1$, and $b=c_p$, while the threshold is set to zero.

Starting with a one dollar investment at the beginning of the sample, our investor decides to hold the stock depending on whether the predicted probability is favorable or not. We compare the cumulative returns from this simple market-timing strategy using predictions from the ordered logit, unordered logit, GARCH, FHS, and the buy and hold strategy for all 29 stocks separately.

\if1\stanislav

{
\newpage

\section{Tables}

Table 1. Estimates of intercepts $\delta _{0,j}$ in the ordered logit
specification $\theta _{t,j}=\delta _{0,j}+x_{t-1,j}^{\prime }\delta \left(
\alpha _{j}\right) $ for the three illustrative stocks. The standard errors are below the point estimates.\bigskip

\begin{equation*}
\label{tab:paramaters1}
\begin{tabular}{ccccccccc}
\hline
$\alpha_j$ & INTC & QCOM & XOM & $\underset{}{}$ & $\alpha_j$ & INTC & QCOM & XOM \\
\cline{1-4}\cline{6-9}
$5\%$ &   $\underset{(0.115)}{-3.046}$ & $\underset{(0.116)}{-2.775}$ & $\underset{(0.118)}{-2.871}$ &  & $95\%$ &   $\underset{(0.155)}{2.898}$ & $\underset{(0.113)}{2.689}$ & $\underset{(0.122)}{3.042}$ \\
$7.5\%$ & $\underset{(0.096)}{-2.504}$ & $\underset{(0.098)}{-2.422}$ & $\underset{(0.102)}{-2.488}$ &  & $92.5\%$ & $\underset{(0.156)}{2.704}$ & $\underset{(0.108)}{2.581}$ & $\underset{(0.098)}{2.579}$ \\
$10\%$ &  $\underset{(0.087)}{-2.195}$ & $\underset{(0.088)}{-2.148}$ & $\underset{(0.092)}{-2.235}$ &  & $90\%$ &   $\underset{(0.095)}{2.026}$ & $\underset{(0.100)}{2.401}$ & $\underset{(0.098)}{2.474}$ \\
$12.5\%$ &$\underset{(0.081)}{-1.953}$ & $\underset{(0.081)}{-1.999}$ & $\underset{(0.088)}{-2.032}$ &  & $87.5\%$ & $\underset{(0.082)}{1.702}$ & $\underset{(0.091)}{2.137}$ & $\underset{(0.089)}{2.105}$  \\
$15\%$ &  $\underset{(0.077)}{-1.800}$ & $\underset{(0.075)}{-1.810}$ & $\underset{(0.083)}{-1.812}$ &  & $85\%$ &   $\underset{(0.079)}{1.666}$ & $\underset{(0.089)}{2.069}$ & $\underset{(0.087)}{1.962}$ \\
$17.5\%$ &$\underset{(0.074)}{-1.672}$ & $\underset{(0.071)}{-1.684}$ & $\underset{(0.082)}{-1.679}$ &  & $82.5\%$ & $\underset{(0.075)}{1.455}$ & $\underset{(0.085)}{1.957}$ & $\underset{(0.083)}{1.751}$ \\
$20\%$ &  $\underset{(0.072)}{-1.489}$ & $\underset{(0.068)}{-1.546}$ & $\underset{(0.082)}{-1.575}$ &  & $80\%$ &   $\underset{(0.072)}{1.338}$ & $\underset{(0.080)}{1.719}$ & $\underset{(0.084)}{1.645}$ \\
$22.5\%$ &$\underset{(0.070)}{-1.325}$ & $\underset{(0.065)}{-1.397}$ & $\underset{(0.082)}{-1.454}$ &  & $77.5\%$ & $\underset{(0.070)}{1.214}$ & $\underset{(0.076)}{1.528}$ & $\underset{(0.083)}{1.492}$ \\
$25\%$ &  $\underset{(0.068)}{-1.151}$ & $\underset{(0.064)}{-1.252}$ & $\underset{(0.081)}{-1.301}$ &  & $75\%$ &   $\underset{(0.066)}{1.049}$ & $\underset{(0.068)}{1.249}$ & $\underset{(0.081)}{1.336}$ \\
$27.5\%$ &$\underset{(0.067)}{-1.020}$ & $\underset{(0.062)}{-1.074}$ & $\underset{(0.081)}{-1.174}$&  & $72.5\%$ &  $\underset{(0.064)}{0.899}$ & $\underset{(0.065)}{1.100}$ & $\underset{(0.081)}{1.171}$ \\
$30\%$ &  $\underset{(0.065)}{-0.859}$ & $\underset{(0.061)}{-0.928}$ & $\underset{(0.080)}{-1.033}$ &  & $70\%$ &   $\underset{(0.063)}{0.828}$ & $\underset{(0.062)}{0.953}$ & $\underset{(0.080)}{1.035}$ \\
$32.5\%$ &$\underset{(0.064)}{-0.711}$ & $\underset{(0.060)}{-0.781}$ & $\underset{(0.080)}{-0.919}$ &  & $67.5\%$ & $\underset{(0.063)}{0.764}$ & $\underset{(0.061)}{0.831}$ & $\underset{(0.080)}{0.873}$ \\
$35\%$ &  $\underset{(0.063)}{-0.631}$ & $\underset{(0.059)}{-0.633}$ & $\underset{(0.080)}{-0.813}$ &  & $65\%$ &   $\underset{(0.062)}{0.650}$ & $\underset{(0.060)}{0.707}$ & $\underset{(0.079)}{0.678}$ \\
$37.5\%$ &$\underset{(0.062)}{-0.477}$ & $\underset{(0.059)}{-0.542}$ & $\underset{(0.080)}{-0.690}$ &  & $62.5\%$ & $\underset{(0.061)}{0.539}$ & $\underset{(0.060)}{0.622}$ & $\underset{(0.078)}{0.531}$ \\
$40\%$ &  $\underset{(0.062)}{-0.386}$ & $\underset{(0.059)}{-0.465}$ & $\underset{(0.080)}{-0.595}$ &  & $60\%$ &   $\underset{(0.061)}{0.425}$ & $\underset{(0.059)}{0.501}$ & $\underset{(0.079)}{0.393}$\\
$42.5\%$ &$\underset{(0.062)}{-0.286}$ & $\underset{(0.059)}{-0.344}$ & $\underset{(0.080)}{-0.499}$ &  & $57.5\%$ & $\underset{(0.061)}{0.334}$ & $\underset{(0.059)}{0.354}$ & $\underset{(0.079)}{0.227}$ \\
$45\%$ &  $\underset{(0.061)}{-0.223}$ & $\underset{(0.059)}{-0.280}$ & $\underset{(0.080)}{-0.345}$ &  & $55\%$ &   $\underset{(0.061)}{0.226}$ & $\underset{(0.059)}{0.253}$ & $\underset{(0.079)}{0.109}$ \\
$47.5\%$ &$\underset{(0.062)}{-0.098}$ & $\underset{(0.059)}{-0.134}$ & $\underset{(0.079)}{-0.205}$ &  & $52.5\%$ & $\underset{(0.061)}{0.096}$ & $\underset{(0.059)}{0.144}$ & $\underset{(0.079)}{0.023}$  \\
\cline{6-9}

$0.50\%$ & $\underset{(0.062)}{-0.004}$ & $\underset{(0.059)}{-0.014}$ & $\underset{(0.079)}{-0.117}$ &  &

 & $\underset{}{}$ &  &  \\ \cline{1-4}
\end{tabular}%
\end{equation*}%
\bigskip

\newpage

Table 2. Estimates of slope coefficients $\kappa _{i,\ell }$ in the ordered
logit specification $\delta _{\ell }\left( \alpha _{j}\right) =\kappa
_{0,\ell }+\sum_{i=1}^{q_{\ell }}2^{i}(\alpha _{j}-0.5)^{i}\cdot \kappa
_{i,\ell }$ for the three illustrative stocks. The standard errors are below the point estimates.\bigskip

\begin{equation*}
\begin{tabular}{cccccccccc}
\hline
Coefficient & INTC & QCOM & XOM & $\underset{}{}$ &  & Coefficient & INTC & QCOM & XOM \\
\cline{1-4}\cline{7-10}
$\kappa_{0,1}$ & $\underset{(0.018)}{-0.003}$ & $\underset{(0.015)}{-0.169}$ & $\underset{(0.011)}{-0.053}$ &  &  & $\kappa_{0,2}$ & $\underset{(3.93)}{-3.33}$ & $\underset{(4.001)}{-0.285}$ & $\underset{(8.104)}{0.109}$ \\
$\kappa_{1,1}$ & $\underset{(0.044)}{-0.035}$ & $\underset{(0.035)}{-0.143}$ & $\underset{(0.043)}{-0.117}$ &  &  & $\kappa_{1,2}$ & $\underset{(5.79)}{4.92}$ & $\underset{(3.08)}{-15.06}$ & $\underset{(5.71)}{-17.52}$ \\
$\kappa_{2,1}$ & $\underset{(0.084)}{0.085}$ & $\underset{(0.092)}{0.542}$ & $\underset{(0.076)}{0.052}$ &  &  & $\kappa_{2,2}$ &    $\underset{(4.72)}{12.13}$ & $\underset{(5.61)}{-7.16}$ & $\underset{(7.83)}{-16.85}$ \\
\cline{1-4}
&  &  &  &  &  & 		$\kappa _{3,2}$ &$\underset{(7.32)}{-3.86}$ & $\underset{(7.524)}{25.34}$ & $\underset{(9.23)}{25.98}$ \\ \cline{7-7}\cline{7-10}
\end{tabular}%
\label{tab:paramaters2}
\end{equation*}%
\bigskip \bigskip

Table 3. Mean and median return-volatility characteristics from five
trading strategies for all 29 stocks.\bigskip

\begin{equation*}
\begin{tabular}{lccccccc}
\hline
& \multicolumn{3}{c}{mean} &  & \multicolumn{3}{c}{median} \\
\cline{2-4}\cline{6-8}
Method & \ \ return\ \  & volatility & \ \ Sharpe\ \  &  & \ \ return\ \  & volatility & \ \ Sharpe\ \  \\
\cline{1-4}\cline{6-8}
Ordered Logit 	& 1.296 & 0.159 & 0.381 & & 0.789 & 0.161 & 0.322 \\
Separate Logits & 1.088 & 0.159 & 0.289 & & 0.663 & 0.155 & 0.289\\
Market 			& 0.806 & 0.244 & 0.102 & & 0.447 & 0.226 & 0.219 \\
GARCH 			& 0.768 & 0.202 & 0.169 & & 0.414 & 0.182 & 0.277 \\
FHS 			& 0.791 & 0.193 & 0.162 & & 0.428 & 0.184 & 0.225 \\ \hline
\end{tabular}%
\label{tab:perf}
\end{equation*}

\newpage

\section{Figures: Parameter Estimates}

\begin{figure}[!h]
\centering
\includegraphics[width=4.8in]{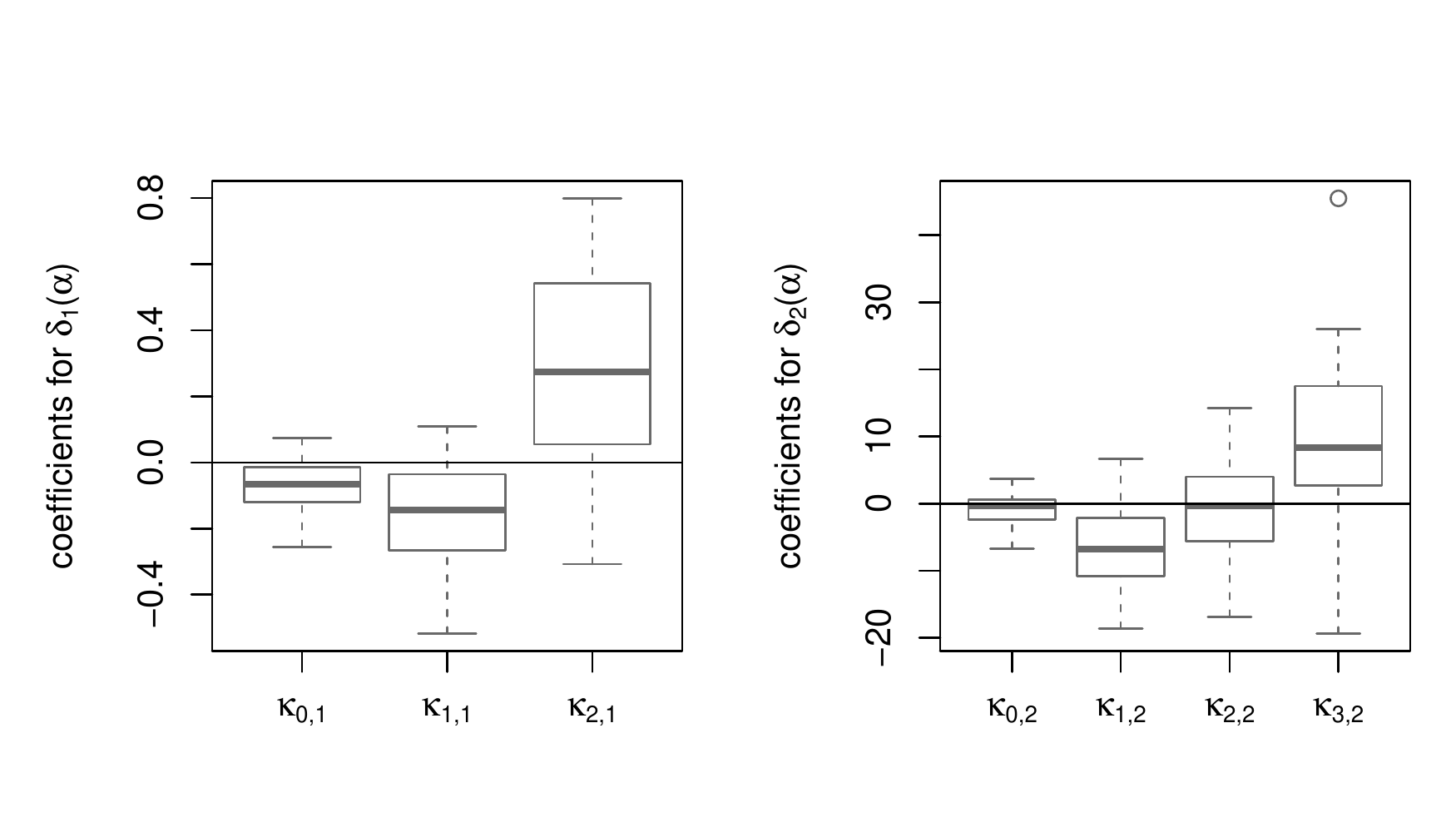}
\caption{Parameter estimates: ordered logit parameters estimated for all 29 stocks shown by box-and-whisker plots.}
\label{fig:pars1}
\end{figure}

\begin{figure}[!h]
\centering
\includegraphics[width=\textwidth]{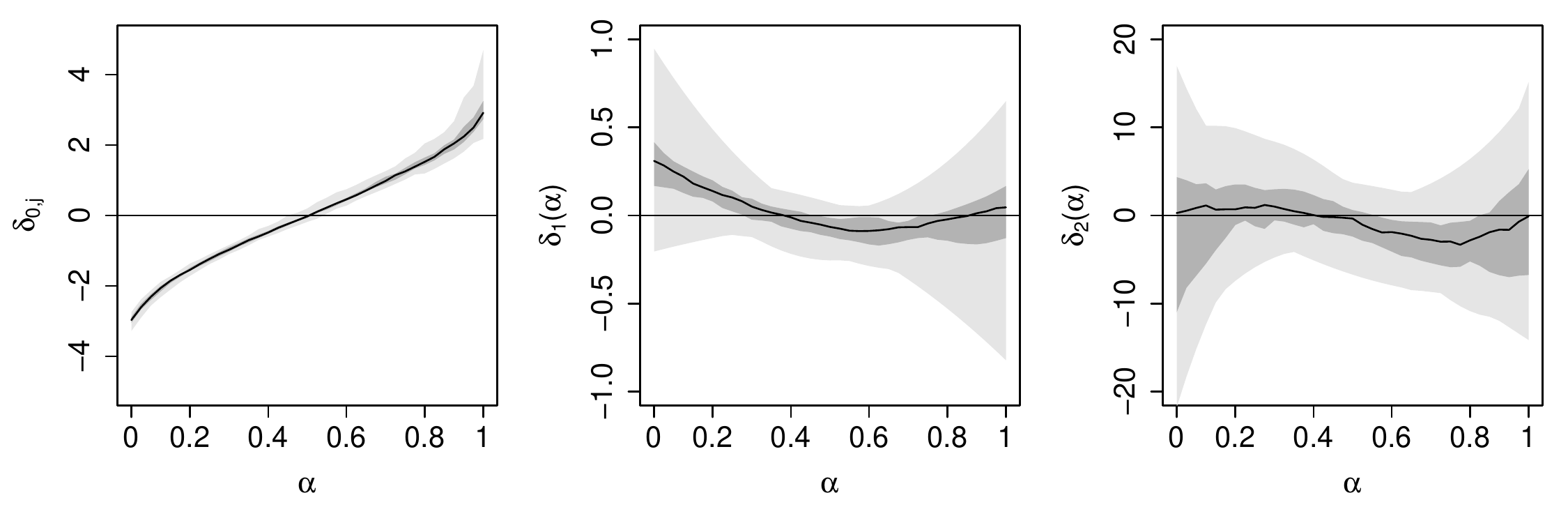}
\caption{Parameter estimates: coefficient functions implied by parameters estimated for all 29 stocks. Minimum and maximum values are shown as a light grey area, 50\% of the distribution as a grey area, and the median as a black line.}
\label{fig:pars2}
\end{figure}

\clearpage
\newpage
\section{Figures: Conditional CDF}
\vspace{2cm}

\begin{figure}[!h]
\centering
\includegraphics[width=0.75\textwidth]{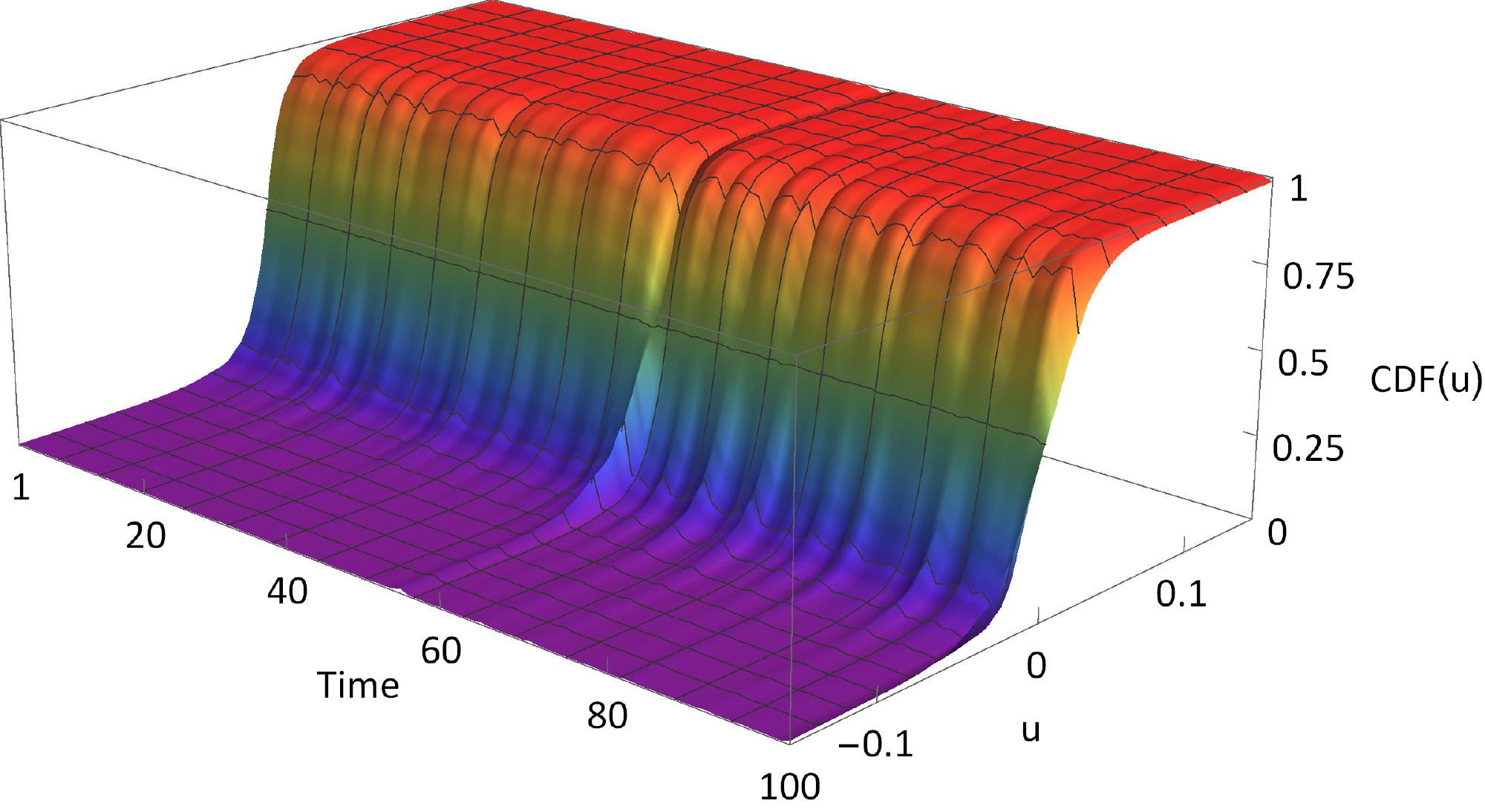}
\par\bigskip\bigskip
\caption{Fragment of interpolated estimated conditional CDF of returns for one of stocks.}
\label{fig:CDFplot}
\end{figure}

\clearpage
\newpage
\section{Figures: Statistical Evaluation}

\begin{figure}[!h]
\centering
\includegraphics[width=0.45 \textwidth]{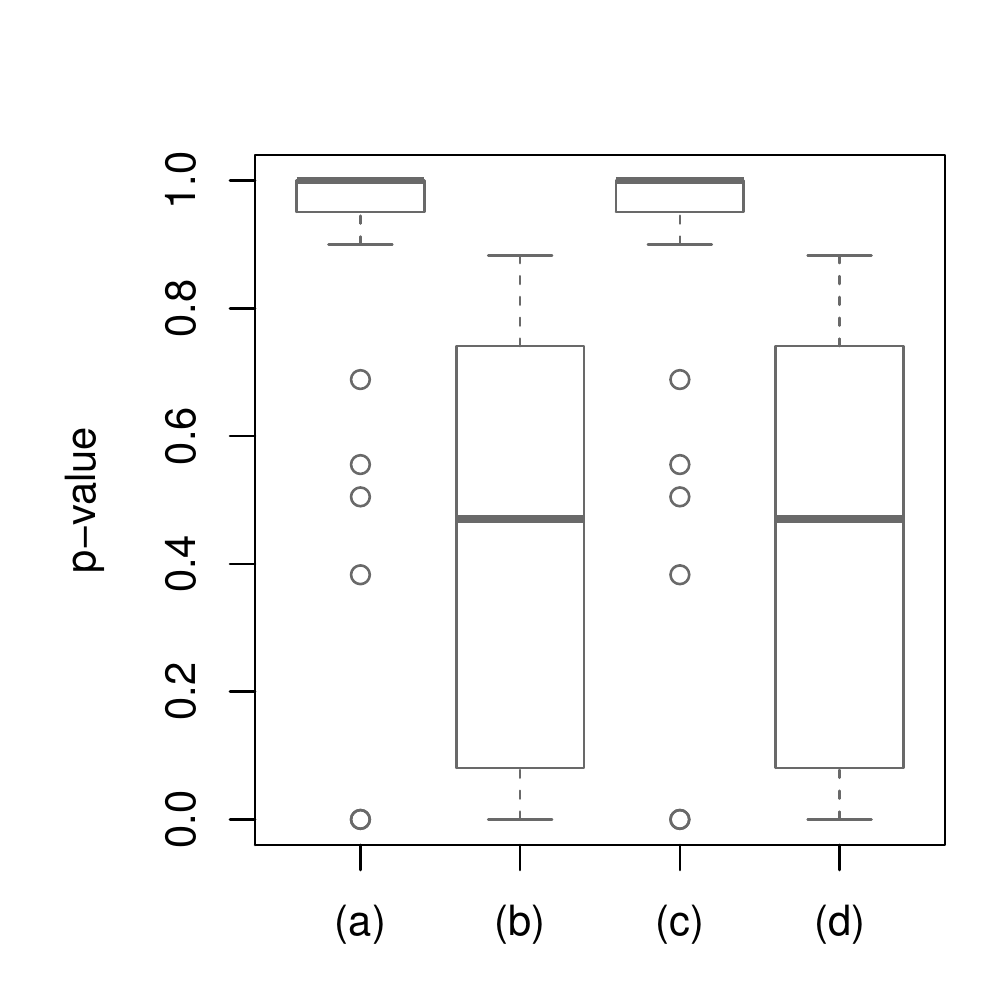}
\caption{p-values from Gonzalez-Rivera and Sun (2015) tests for the ordered logit model shown by box-and-whisker plots for all 29 stocks. The four test specifications are shown with $\alpha = (0.25,0.5,0.75)'$ (a) contour-aggregated, (b) lag-aggregated, with $\alpha = (0.05,0.1,...,0.9,0.95)'$ (c) contour-aggregated, (d) lag-aggregated.}
\label{fig:testplot}
\end{figure}

\begin{figure}[!h]
\centering
\includegraphics[width=0.9\textwidth]{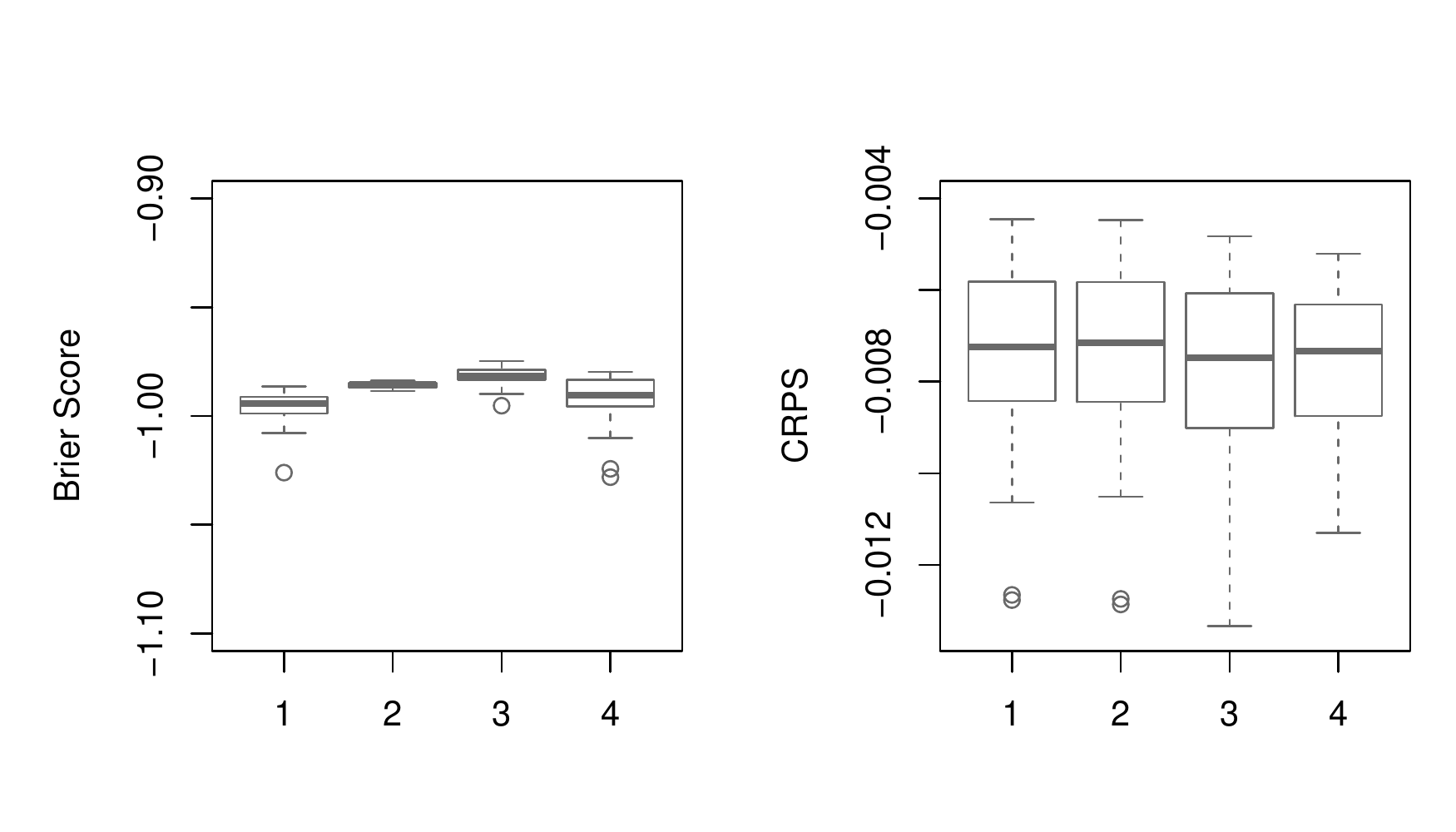}
\caption{Brier scores and CRPS for ordered logit (1), separate logits (2), GARCH (3), and FHS (4), shown by box-and-whisker plots for all 29 stocks. Larger values of the scores are preferred to their smaller values.}
\label{fig:scoresplot}
\end{figure}

\clearpage
\section{Figures: Economic Evaluation}

\begin{figure}[!h]
\centering
\includegraphics[width=\textwidth]{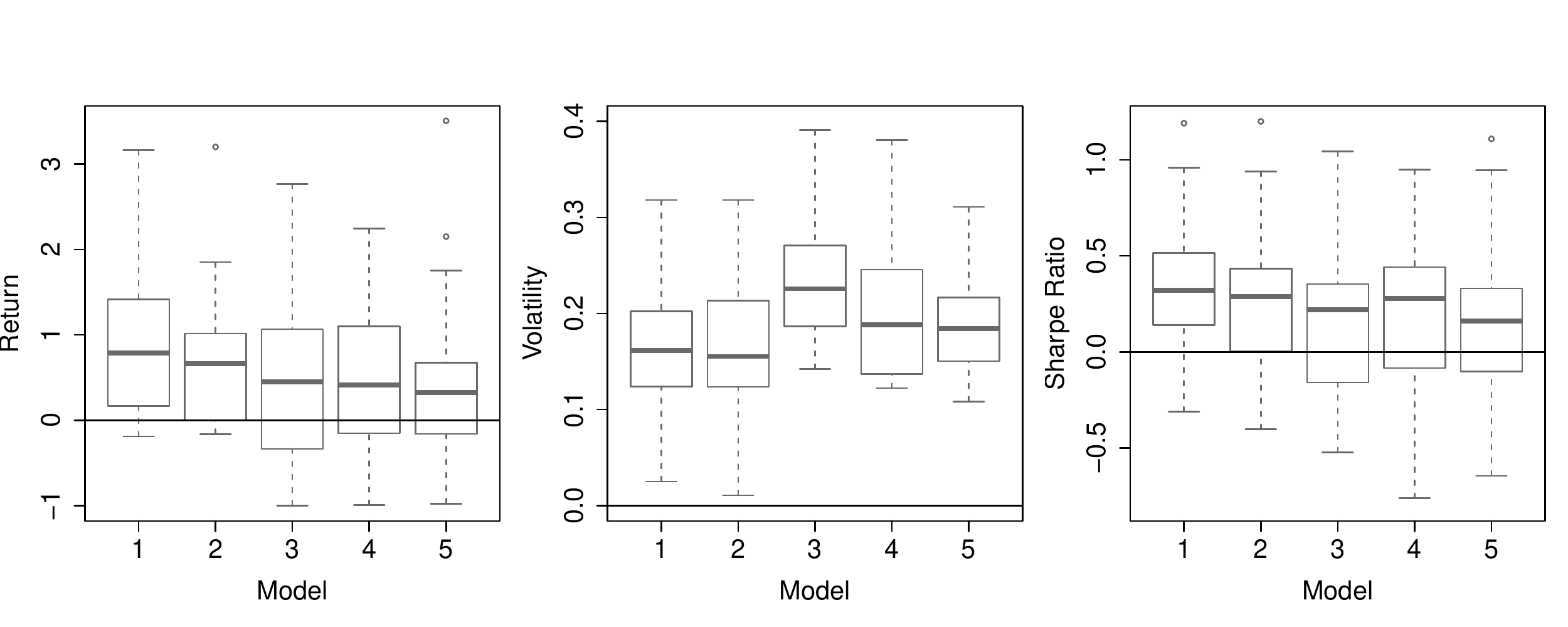}
\caption{Performance of ordered logit (1), separate logits (2), market benchmark (3), GARCH (4), and FHS (5) models. Return, volatility and Sharpe ratio for all 29 stocks are shown by box-and-whisker plots.}
\label{fig:tradingplot}
\end{figure}

\begin{sidewaysfigure}[!ht]
\centering
\includegraphics[width=3in]{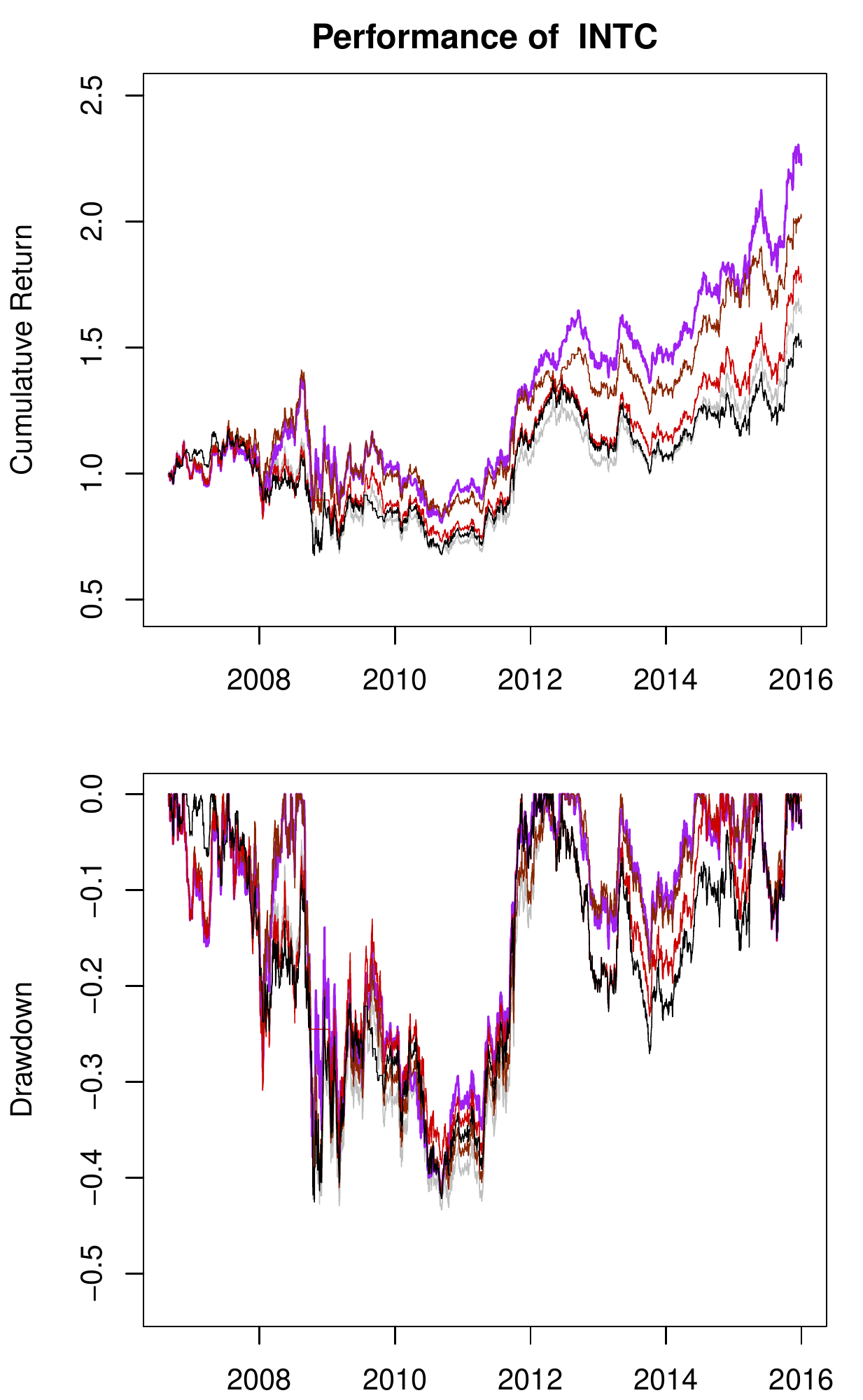}
\includegraphics[width=3in]{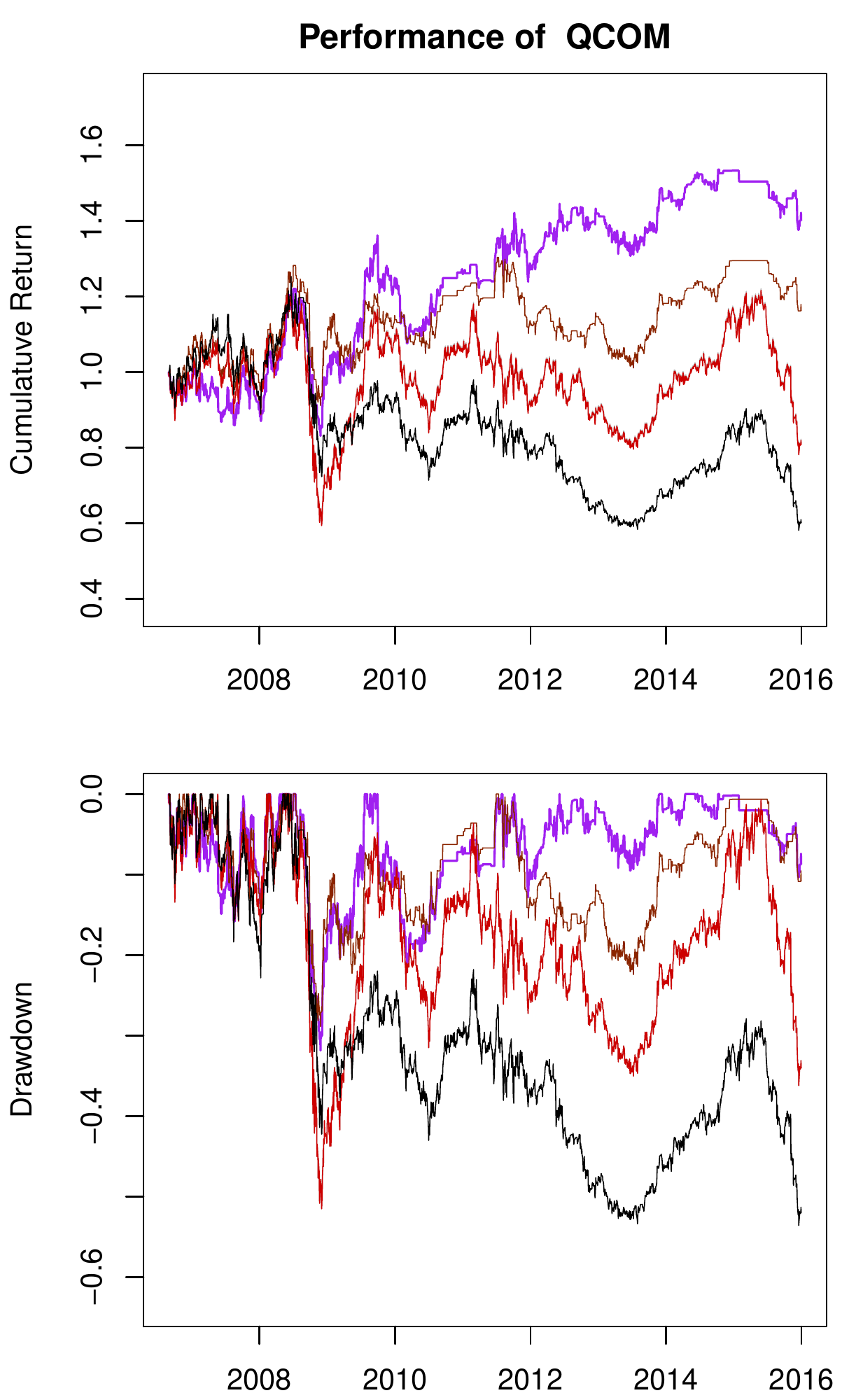}
\includegraphics[width=3in]{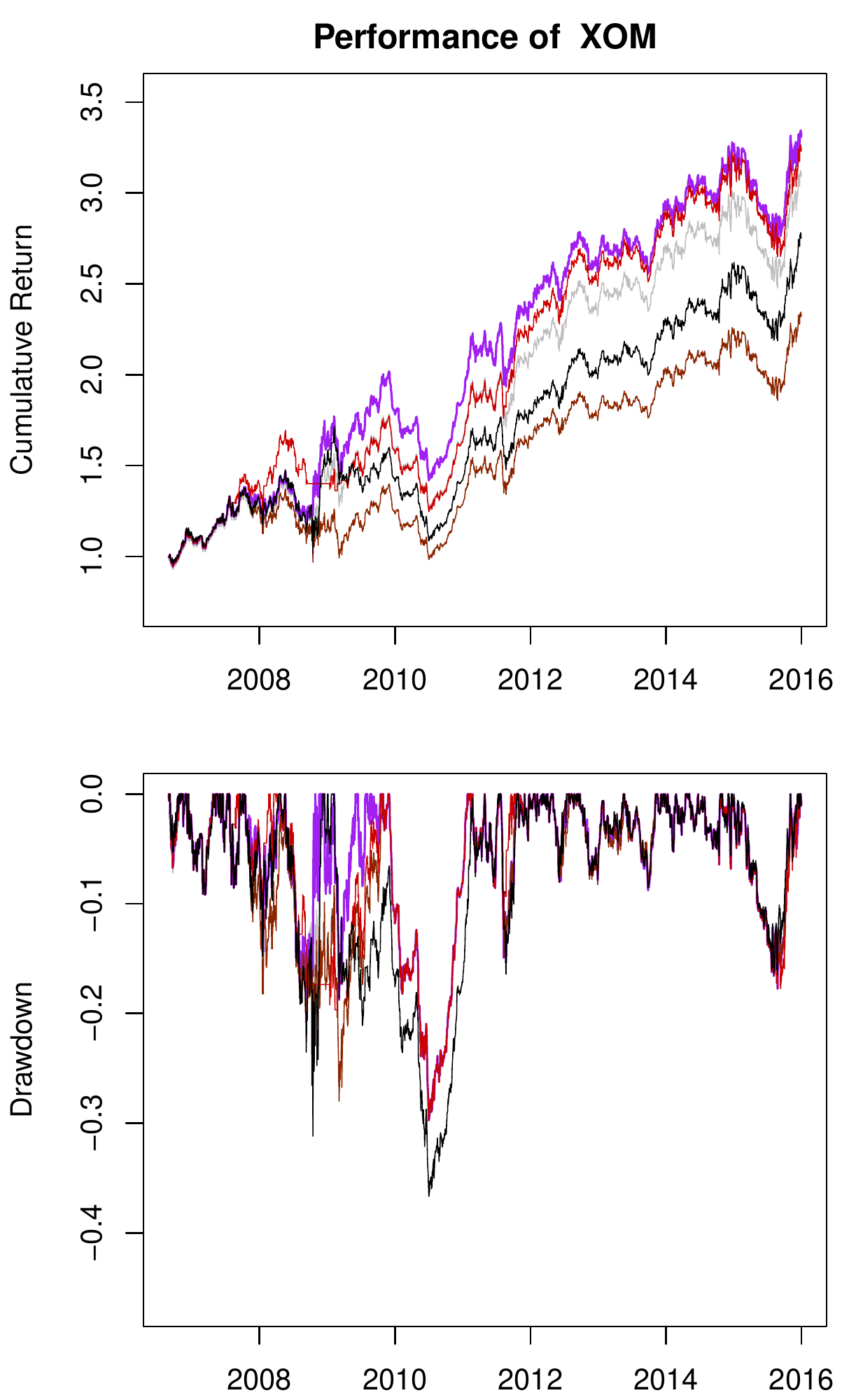}
\caption{Performance: cumulative return and drawdown for three typical stocks. Trading strategy based on probability predictions from ordered logit model in purple ({\color{purple}{ $\boldsymbol{ -}$ }}), separate logits in bordeaux ({\color{bordeaux}{ $\boldsymbol{-}$ }}), GARCH in red ({\color{red}{ $\boldsymbol{-}$ }}), FHS in black ({\color{black}{ $\boldsymbol{-}$ }}), and benchmark buy and hold in grey ({\color{grey}{ $\boldsymbol{-}$ }}).}
\label{fig:exthree}
\end{sidewaysfigure}

\begin{figure}[!h]
\centering
\includegraphics[width=0.8\textwidth]{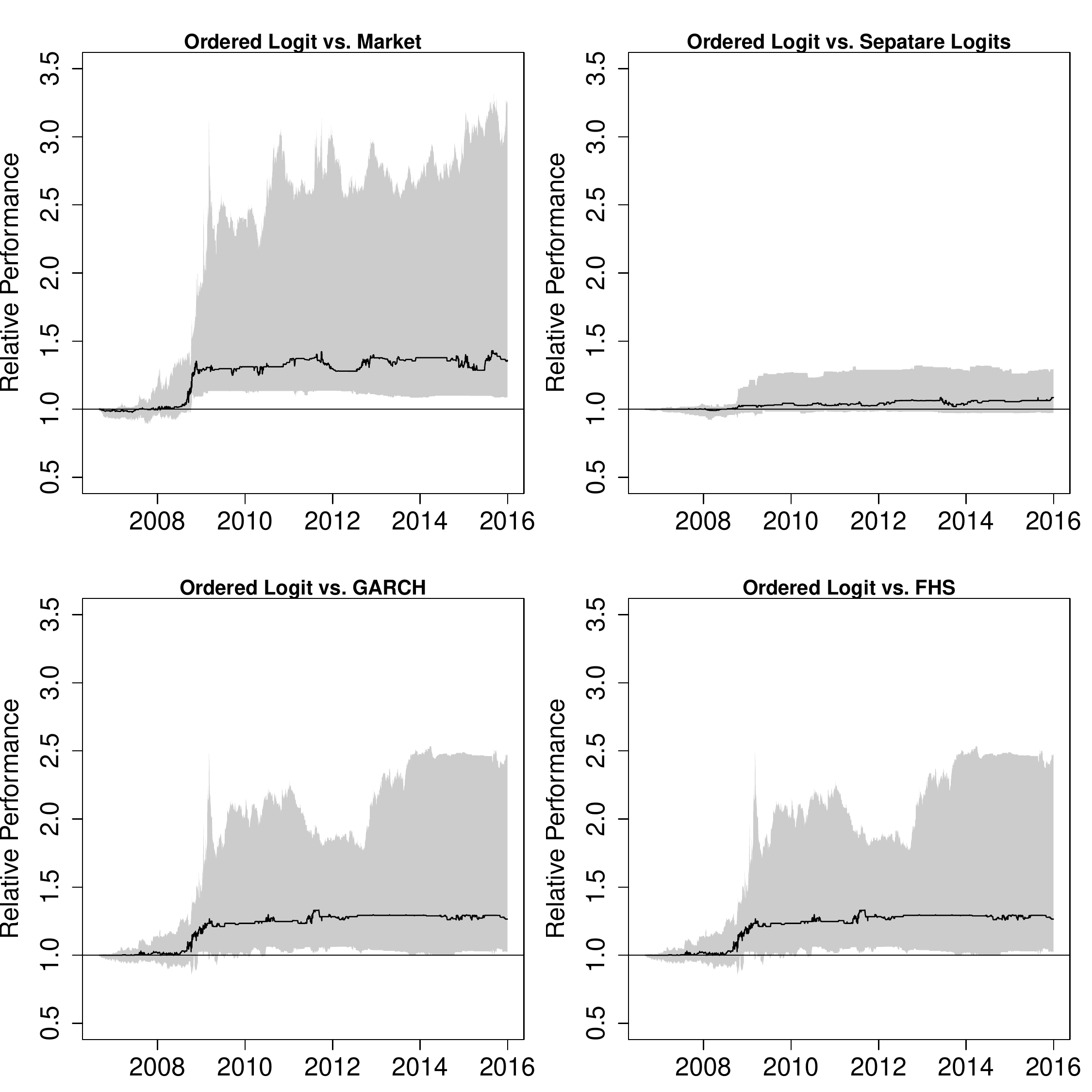}
\caption{Relative performance: trading strategy based on probability predictions from the ordered logit model relative to separate logits (top right) as well as benchmark market (top left), GARCH (bottom left), and FHS (bottom right). The median value from all 29 stocks is black line surrounded by 90\% of the distribution in grey area. Note that the value of one shows equal performance for both of the compared strategies.}
\label{fig:relative}
\end{figure}

\clearpage
\newpage
\section{Figures: Sensitivity to Polynomial Orders}

\begin{figure}[!h]
\centering
\includegraphics[width=0.45 \textwidth]{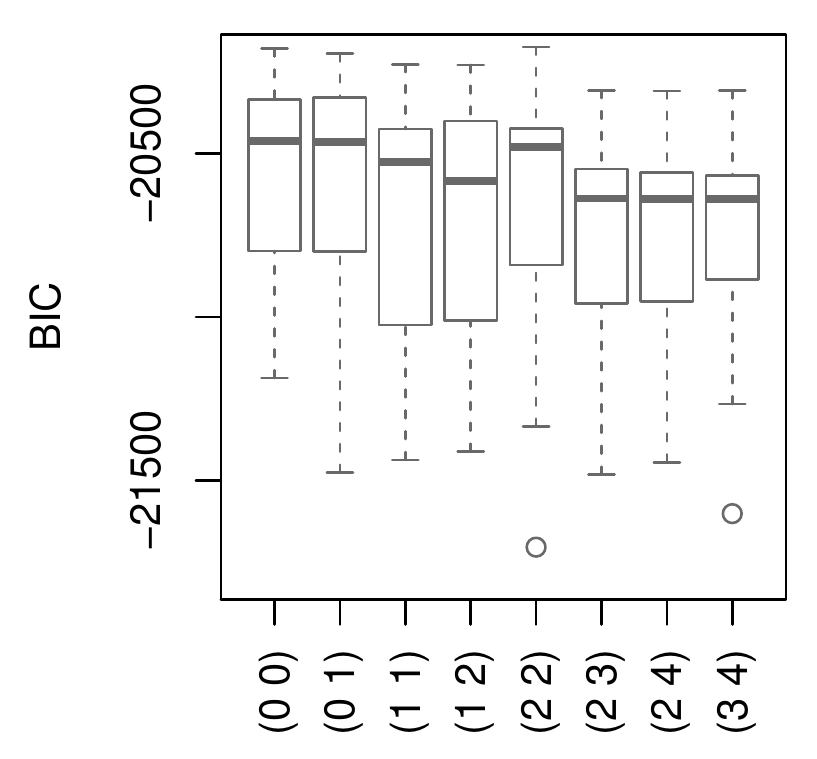}
\caption{Bayesian information criterion for the ordered logit model with different polynomial orders $(q_1\,q_2)$ on the $x$-axis shown by box-and-whisker plots for all 29 stocks.}
\label{fig:sensitivityplot}
\end{figure}

} \fi

\end{document}